\documentclass[preprint]{aastex}

\usepackage{epsf}
\newcommand{\Ts}{\mbox{$<T_{\mathrm s}>$}}
\newcommand{\NHI}{\mbox{$N_\mathrm{HI}$}}
\newcommand{\HI}{\mbox{H\,{\sc i}}}

\begin{document}

\title {Detection of Warm and Cold Phases of the Neutral ISM in a Damped
  Ly$\alpha$ Absorber}

\author{W.M. Lane, F.H. Briggs}

\affil{Kapteyn Astronomical Institute, Postbus 800, NL-9700 AV
  Groningen, The Netherlands.}
\email{wlane@astro.rug.nl, fbriggs@astro.rug.nl}

\and
\author{A. Smette\altaffilmark{1}}
\email{asmette@band3.gsfc.nasa.gov}
\affil{Laboratory for Astronomy and Space Physics, NASA
    Goddard Space Flight Center, Code 681, Greenbelt, MD 20771.}  
\altaffiltext{1}{National Optical Astronomy Observatories, 950 North
    Cherry Avenue, Tucson, AZ 85726.}

\begin{abstract}  
  
  We present a detailed study of the \HI\ 21cm absorption system at
  $z=0.0912$ towards the radio quasar B0738+313.  The uncommonly
  narrow main absorption line and weak secondary line are resolved for
  the first time and have FWHM velocities of $\Delta v_1 = 3.7$ and
  $\Delta v_2 = 2.2$ km s$^{-1}$ (main and secondary component
  respectively).  In addition we find it necessary to add a third,
  broader shallow component to obtain a good fit to the spectrum.
  Although the harmonic mean spin temperature calculated by
  comparison of the 21cm lines to the damped Ly$\alpha$ line is \Ts $=
  775 \pm 100$ K, the thermal kinetic temperatures of the two narrow
  components, calculated from their widths, are much lower:
  $T_{\mathrm k} \leq 297 \pm 3$ and $\leq 103 \pm 10$ K respectively.
  This is the first case of a redshifted absorption system for which
  $T_{\mathrm k}$ is measured to be {\it less} than \Ts.  We discuss
  this result in the context of a two phase gas model, in which the
  damped Ly$\alpha$ gas is sensitive to a significant neutral column
  density of warm phase gas as well as the cold phase gas of the
  narrow 21cm lines.  The third component is interpreted as
  representing the warm phase gas with with $T_{\mathrm k} \leq 5050
  \pm 950$ K.  The combined column density of the three 21cm
  components is approximately equal to that derived from fits to the
  damped Ly$\alpha$ line.

\end {abstract}


\clearpage

\section{INTRODUCTION}

Damped Ly$\alpha$ (DLA) and \HI\ 21cm absorption lines are the
spectral signatures of systems with high neutral hydrogen column
densities.  Such systems are the major contributors to the mass
density of neutral gas at high redshifts ($z \approx 3$).  Despite the
success of several surveys to identify DLA absorption lines in both
the optical (cf.  Wolfe et al. 1986; Lanzetta et al. 1991; Wolfe et
al. 1995; Storrie-Lombardi et al. 1996) and the UV regimes (cf.
Lanzetta, Wolfe, \& Turnshek 1995; Rao \& Turnshek 1999), our understanding
of the morphology and evolution of the host galaxies in which they
arise remains poor.  Until recently, most of the known DLA systems
were at high redshift, making identification and study of the host
galaxy difficult.

The profiles of associated metal lines in these systems are consistent
with the hypothesis that DLAs arise in rapidly rotating large disks
with substantial vertical scale heights (Prochaska \& Wolfe 1997,
1998), which fits well with the standard paradigm that DLAs arise in
luminous disk galaxies or their progenitors (cf. Wolfe et al. 1995).
However, there is a growing body of evidence in both theory and
observations that this model might not be applicable to all DLA
absorbers.  One theory suggests that the interactions of relatively
small systems or even proto-galactic clumps might account for the line
profiles and neutral column densities of these absorbers (Rauch,
Haehnelt, \& Steinmetz 1997; Haehnelt, Steinmetz, \& Rauch 1998).  A
number of spectral and imaging studies of low and moderate redshift
DLAs have identified a variety of host galaxies with moderate to high
luminosities, including compact objects and amorphous LSB galaxies as
well as spiral disks (Steidel et al.  1997; Le Brun et al. 1997; Rao
\& Turnshek 1998).

Just as the morphology of DLA systems is poorly understood, so are
their physical conditions, and in particular the temperature of the
gas in which the absorption arises.  Under usual conditions in the
ISM, the excitation, or spin temperature, $T_{\mathrm s}$, describing
the hyperfine level populations in the ground state of neutral
hydrogen is coupled to its thermal kinetic temperature, T$_{\mathrm
  k}$.  A harmonic mean spin temperature, \Ts, of all of the neutral
clouds on a given sightline can be determined by a comparison of \HI\
21cm absorption and an inferred 21cm emission profile for the same
sightline, or by comparison of the 21cm absorption line parameters to
a column density from the DLA absorption feature on the same
sightline.  However the values of \Ts\ derived in this manner for
redshifted DLA/\HI\ 21cm absorbers are consistently higher than those
found in clouds of similar optical depth in the Galaxy (cf. Lane et
al. 1998; Carilli et al. 1996).

The two DLA absorption systems identified in the UV spectrum of the
$z_\mathrm{em} = 0.630$ QSO B0738+313 (OI 363), are both examples of
low-redshift absorbers with no corresponding luminous disk galaxy (Rao
\& Turnshek 1998).  They are both \HI\ 21cm absorbers, and both have
very narrow 21cm lines (Lane et al. 1998; Chengalur \& Kanekar 1999).
The first absorber is the lowest redshift system for which the DLA
line has been observed (note that the ``DLA System'' at $v \approx
1170$ km s$^{-1}$ reported by Miller, Knezek, \& Bregman (1999) was
not observed in the Lyman lines, and a measurement of x-ray absorption
was used to infer the neutral HI column density).  It has a redshift
$z_1 = 0.0912$ and a neutral hydrogen column density of \NHI$ = 1.5
\pm 0.2 \times 10^{21}$ cm$^{-2}$.  It was discovered serendipitously
in a HST-FOS spectrum taken to observe the Ly-$\alpha$ line associated
with the previously known metal-line absorption system at $z_2=0.2212$
on the same sightline.  This second system is also damped, with a
slightly lower column density of \NHI$ = 7.9 \pm 1.4 \times 10^{20}$
cm$^{-2}$.  In ground-based optical images only one candidate host
galaxy is visible near the QSO sightline.  Based on its optical
brightness and the assumption that it is a moderatly luminous galaxy,
this candidate absorber has been associated with the higher redshift
system, although it has no confirming spectroscopic redshift (LeBrun
et al. 1993).  The other absorption system must arise in either an LSB
galaxy or a small galaxy which falls under the PSF of the QSO (Rao \&
Turnshek 1998).

Here we present new information on the lower redshift ($z_1=0.0912$)
system on the B0738+313 sightline.  Very little is known about the
associated metal lines, because the wavelength coverage of the only
published optical spectrum (Boiss{\' e} et al. 1992) does not include
the expected wavelengths of any strong lines belonging to this system,
while the HST-FOS spectrum (Rao \& Turnshek 1998) has insufficient
signal to noise and resolution to allow the detection of small
equivalent width lines.  We present WHT/ISIS\footnote{The William
  Herschel Telescope (WHT) is operated on the island of La Palma by
  the Isaac Newton Group in the Spanish Observatorio del Roque de los
  Muchachos of the Instituto de Astrofisica de Canarias} observations
of CaII H and K, the only metal lines observed so far in this system.
We discuss observations of the \HI\ 21cm absorption made with the VLA
(Very Large Array)\footnote{operated by The National Radio Astronomy
  Observatory, a facility of the National Science Foundation operated
  under cooperative agreement by Associated Universities, Inc.}, and
the Arecibo Radio Telescope\footnote{The Arecibo Observatory is part
  of the National Astronomy and Ionosphere Center, which is operated
  by Cornell University under a cooperative agreement with the
  National Science Foundation}, and the estimates of temperature which
we derive from them.  We also present Very Long Baseline Array
(VLBA)$^{5}$ data which was used to investigate the line
characteristics and gas covering factor on parsec, or cloud scales.
Because the VLBA achieves subarcsecond resolution, data from these
observations provide a fair comparison to the optical data.

\section{OBSERVATIONS}
\subsection{WHT/ISIS} 

The initial redshift determination of the $z=0.09$ absorber on the
B0738+313 sightline had an unreasonably large uncertainty due largely
to the difficulties in fitting damped absorption profiles, and the low
signal to noise of the HST detection spectrum.  Because an accurate
redshift facilitates the search for associated 21cm absorption,
observations were made of strong metal lines which fall in the optical
window at this redshift.  Spectra were obtained using the WHT+ISIS in
service mode on the night of 1997 November 9-10, by R.  Rutten.  The
detector was the 2148x4200 EEV10 CCD.  We used the R600B grating with
a central wavelength of 4300 \AA, so that the spectrum covers the
range $\lambda\lambda$ 3420\AA -- 5200\AA.  The slit was oriented
along the parallactic angle; its width was 1.00" leading to a spectral
resolution of $\sim 1$\AA.  Two exposures of 900s and one exposure of
1800s were obtained, for a total integration time of 1 hour.  The data
were reduced with standard MIDAS procedures.

An excerpt of the resulting spectrum is presented in Fig. 1. It shows
the Ca {\sc ii} H and K absorption lines with rest equivalent widths
of $0.11\pm0.02$ and $0.15\pm0.02$, respectively, at a mean redshift
$z = 0.09103\pm0.00007$, where the uncertainty represents mainly the
RMS of the wavelength calibration.  No significant absorption is seen
at the expected positions for Fe {\sc i} $\lambda$3720, Al {\sc i}
$\lambda$3945, or Ca {\sc i} $\lambda$4227.  For completeness, we note
that the large peak near 4570 \AA\ is Mg {\sc ii} emission from the QSO
(z=0.63), and the region between 4050 and 4320 \AA\ contains a complex
of emission lines attributed to Fe {\sc ii} $\lambda$2500.

\subsection{VLA}

The VLA was used in D-array configuration for 4 hours on 1998 January
3 to look for associated \HI\ 21cm absorption at the redshift of the
CaII lines.  The sources 3C147 and B0735+331 were used to calibrate
the amplitude and passband.  The observations were made in 2
polarizations, with a bandwidth of 3.125 MHz and 128 channels for a
channel separation of 5.6 km/s. The data were reduced using standard
routines in AIPS.  An unresolved 21cm absorption line was observed at
a redshift of $z=0.09118$, with an optical depth $\tau = 0.08$ and a
FWHM velocity $\Delta v = 13.4$ km s$^{-1}$.  By comparing the 21cm
line characteristics with the column density from the DLA measurement
(see equations in section 5), a harmonic mean spin temperature of \Ts
$= 775 \pm 100$ K was derived.  This temperature represents a column
density weighted harmonic mean of the spin temperatures of all of the
neutral gas clouds along the sightline.  The derived harmonic mean
spin temperature is typical of values found in most other redshifted
21cm/DLA absorbers, and is considerably higher than typical values at
similar optical depths in the Milky Way (Lane et al. 1998 and
references therein).

\subsection{VLBA Observations}

The $z=0.0912$ line was observed on 1998 October 8 for just under 6
hours, using the VLBA array, as part of a project that observed both
21cm lines on this QSO sightline simultaneously.  All 10 VLBA stations
participated in the experiment, but due to technical problems, data
from Kitt Peak and Los Alamos were not useful.  Pie Town had a clock
error which was corrected halfway through the experiment.  The compact
source B0742+103 was observed for calibration purposes.  The data were
reduced using standard routines in the AIPS data reduction package.
The VLBA system has a poor response at 1163 MHz, so the observations
of the $z=0.2212$ system provided little useful information.

At 1302 MHz the QSO center is lightly resolved into a core and an extended
component, as shown in Figure 3, and all of the ``core'' flux from
unresolved observations is recovered.  The symmetric, weak, extended
lobes, which lie at roughly 30$''$ to the north and south of the
quasar (Murphy, Brown, \& Perley 1993), have been resolved away.
Spectra from three locations across the QSO have been extracted.  The
upper spectrum is offset by one synthesized beam (FWHP) to the north;
the lower offset is one beam to the south, and two beams to the east,
corresponding to a position along the extension of the QSO.  Although
nearly 65\% of the continuum level in the upper spectrum is due to
contaminating signal from the center sightline, the lower position is
nearly completely independent and is unlikely to be contaminated at a
detectable level.  The total shift in position on the sky between the
center and lowermost sightline is 13.2 mas, which corresponds to 20
$h^{-1}_{75}$ pc at the redshift of the absorber.

\subsection{Arecibo}
  
The Arecibo Radio Telescope was used to make three one-hour on source
integrations of the 1665 and 1667 MHz OH lines in the two absorption
systems on the B0738+313 sightline.  Given the narrowness of the 21cm
absorption features at both redshifts (Lane et al. 1998; Chengalur \&
Kanekar 1999), we knew that the absorbing clouds must be very cold and
therefore might contain detectable molecular gas.  The observations
were carried out by K. O'Neill on 1998 November 30 and 1998 December
1\&3.  Using the post-upgrade system, the L-Band wide receiver, and
the new autocorrelation spectrometer, it was possible to observe both
OH lines at the redshifts of both DLA systems simultaneously by
placing 4 bandpass subcorrelators at appropriate frequencies.  A
bandwidth of 3.125 MHz and 1024 channels gave a velocity resolution of
0.7 km s$^{-1}$ in two polarizations. The observations were made in an
on/off ``total power'' mode.  Total power spectra for the calibration
source 3C236 were processed to form gain templates for the spectral
passbands, and these templates were applied to the B0738+313
difference spectra using Analyz software.  The band placed around the
OH 1665 MHz line at $z_1=0.0912$ was corrupted by interference, and
could not be used.  No OH was detected in the other three bands (OH
1667 MHz at $z_1=0.0912$ and both OH 1667 and 1665 MHz at
$z_2=0.2212$), which were Hanning smoothed to a velocity resolution of
$\delta v = 1.4$ km s$^{-1}$ and had a 3$\sigma$ detection limit of
$\tau = 0.0012$.

On 1999 April 24-25, we observed the 21cm line in the $z_1=0.0912$
system for 45 minutes on source using the on/off observation
technique, in two polarizations.  Placing 2048 channels across the
3.125 MHz bandwidth gave a velocity resolution of 0.35 km s$^{-1}$.
The nearby source J0745+317 was observed as a calibrator. After gain
calibration and averaging, a linear baseline was removed from the data
using Analyz to produce the spectrum shown in figure 2.  Two narrow
absorption lines, the principal component observed in the VLA data and
the weak secondary component first seen in the GMRT spectrum of
Chengular \& Kanekar (1999), are clearly resolved and separated from
each other in the Arecibo spectrum.  In fitting the absorption
features we assume that the signal from any 21cm emission in this
system is negligible.  A sensitive WSRT spectrum (W. Lane, in
preparation) has placed a 3$\sigma$ upper limit to the \HI\ mass of
M$_{\mathrm {HI}} \leq 3 \times 10^9$ M$_{\odot}$ for an assumed
velocity spread of 100 km s$^{-1}$ and H$_o = 75$ km s$^{-1}$
Mpc$^{-1}$.  At the velocity resolution of the Arecibo observations,
this limit falls well within the spectral noise.

\section{COLUMN DENSITY AND TEMPERATURE}

The neutral Hydrogen column density in the 21cm line can be calculated using the
standard equation:
\begin{equation}
{\NHI\ = 1.8\times10^{18}~{{T_{\rm s}}\over{f}}~EW_{21} ~ {\rm cm}^{-2}} 
\end{equation}
where $f$ is the fraction of the continuum source covered by the
absorber, T$_s$ is the spin temperature of the gas, and $EW_{21}$ is
the integral of the optical depth over the velocity range of the line
in units of km s$^{-1}$.  For a single line with a Gaussian profile,
$\tau(v)$,
\begin{equation}
{ EW_{21} = 1.06\times\tau_{\rm c}~\Delta v} 
\end{equation}
where $\tau_{\mathrm c}$ is the
peak optical depth of the line at the line center and $\Delta v$ is
the FWHM velocity in km s$^{-1}$.

The covering factor $f$ of the gas in 21cm absorbers is usually
assumed to be $f \equiv 1$ for compact QSOs, based on a variety of
arguments (cf.  Carilli et al. 1996).  VLBI measurements can be used
to refine estimates for $f$ for more complex sources (Briggs \& Wolfe
1983).  Adoption of a value for $f$ allows a calculation of either the
column density, if the temperature is known, or of the temperature if
the neutral column density is known from a DLA measurement.  B0738+313
is a core dominated quasar, with only $\sim 2\%$ of its total flux
found in weak lobes which extend $\approx 30\arcsec$ north and south
of the core at this frequency (W. Lane, in preparation; Murphy et al.
1993).  New WSRT observations (W. Lane, in preparation) find no
evidence for the presence of the main component line against the
extended lobes, and any other absorption is ruled out at 3$\sigma$
optical depths of $\tau_\mathrm{north} \approx 0.03$ and
$\tau_\mathrm{south} \approx 0.08$, for velocity resolution $\delta v
= 4.5$ km s$^{-1}$) over velocities in a range of several hundred km
s$^{-1}$ to either side of the main component redshift.  This implies
that the covering factor of the absorbing gas is $f \leq 0.98$ over
the entire QSO.

Within the errors, the width, depth and redshift of the main
absorption line do not change across the core of the quasar in the
VLBA data.  The size scale and low velocity dispersion of the gas
causing the main absorption feature suggest that it belongs to a
single cold cloud or cloud complex.  More sensitive VLBA measurements
would be necessary to determine if the weak secondary absorption is
also present over this size scale, or is part of a smaller scale
feature.  The constancy of the primary line characteristics over the
slightly resolved core in the VLBA data suggests that it is completely
covered ($f_\mathrm{core} = 1$) by the absorbing gas in that
component, and we assume this is true for the second weaker component
as well.  We therefore adopt the view that the absorbing gas entirely
covers the core but not the extended weak lobes, and that the covering
factor of the gas is $f \approx 0.98$.

The extreme narrowness of the absorption features (as measured in the
Arecibo data) places firm upper limits on the amount of thermal
broadening in the lines, and consequently on any turbulence or bulk
motions of the absorbing gas.  As a result, an upper limit to the
kinetic temperature for each cloud can be found directly from the
widths of the lines, without having to rely on a comparison between
the DLA and 21cm line characteristics to calculate a harmonic mean
spin temperature for the entire ensemble of clouds which lie on the
sightline.  The kinetic temperature of the gas is related to the width
of the absorption line by:
\begin{equation}
 T_{\rm k} \leq {{1.2119 \times 10^2 \Delta v^2}\over{8 \ln 2}} ~~{\rm K}
\end{equation}
where $\Delta v$ is the FWHM velocity measured in km s$^{-1}$.

\section{WARM PHASE GAS}

A simultaneous two component fit to the Arecibo spectrum left large
residuals and had a reduced $\chi^2 = 2.34$.  By adding a third
component to the fit, the reduced $\chi^2 = 1.02$, and the residuals
were all within the noise.  The first or main component has a velocity
width at half maximum (FWHM) of $\Delta v = 3.687 \pm 0.019$ km
s$^{-1}$, and an optical depth of $\tau = 0.2462 \pm 0.0010$.  It lies
at a heliocentric frequency of 1301.6496 MHz, corresponding to $z =
0.09123$.  The second, weaker line is separated from the first by
$\Delta V_\mathrm{offset} = 7.69 \pm 0.04$ km s$^{-1}$.  It has a FWHM
velocity of $\Delta v = 2.18 \pm 0.11$ km s$^{-1}$ and an optical
depth $\tau = 0.0253 \pm 0.0010$.  The third component has an optical
depth $\tau = 0.0063 \pm 0.0008$ and FWHM velocity $\Delta v = 15.2
\pm 1.4$ km s$^{-1}$.  The absorption is shifted by $\Delta
V_\mathrm{offset} = -1.59 \pm 0.66$ km s$^{-1}$ with respect to the
main narrow absorption feature.  Attempts to force the third component
to lie at the position of either of the other two components resulted
in much poorer fits.  Figure 4 shows the Arecibo spectrum after the
fits to the first two components have been removed.  The fit to the
third component has been marked, and the residuals after removing all
three gaussian fits are shown.  Parameters for each of the absorption
components are summarized in Table 1.

Using equation (3), the kinetic temperature is $T_{\mathrm k} \leq 297 \pm
3$ K for the main component and $T_{\mathrm k} \leq 103 \pm 10$ K for
the secondary line.  Both of these temperatures would fall within the
scatter in the Galactic relation for \Ts-$\tau$ (Braun \& Walterbos
1992), unlike the considerably higher temperature \Ts $= 775 \pm 100$
K that was derived earlier.  This is the first redshifted system for
which it can be shown that $T_{\mathrm k}$ is less than \Ts, ie.  that
the kinetic temperature in the individual cold phase gas clouds is
less than the derived harmonic mean spin temperature for all of the
neutral gas on the sightline.  The kinetic temperature of the third
line, derived from the velocity width of the line, is $T_{\mathrm k}
\leq 5046 \pm 953$ K, in reasonable agreement with measurements of
temperature in the WNM of our own Galaxy (Kulkarni \& Heiles 1988;
Carilli, Dwarakanath \& Goss 1998), where typical temperatures fall in
the range 5000-8000 K.

For a given cloud, $T_{\mathrm s} \approx$ $T_{\mathrm k}$ under usual
conditions found in the ISM (Kulkarni \& Heiles 1988). For the two
cold absorption components, we set $T_{\mathrm s} =$ $T_{\mathrm k}$
and $f= 0.98$ (because the Arecibo beam covers both the core and the
extended lobes of the quasar), and calculate the column density for
each 21cm line component from equations (1) and (2).  Adding the two
together, and bearing in mind that our values for $T_{\mathrm k}$ are
upper limits, we find a total column of \NHI$_\mathrm{21cm} \leq 5.4
\pm 0.1 \times 10^{20}$ cm$^{-2}$ in the narrow absorption features.
This is approximately one third of the measured \HI\ column density in
the DLA line: \NHI$_\mathrm{DLA} = 1.5 \pm 0.2 \times 10^{21}$
cm$^{-2}$ (Rao \& Turnshek, 1998).

The calculated column density in the warm component is \NHI\ $ \leq
9.4 \pm 2.3 \times 10^{20} $ cm$^{-2}$ for a covering factor of $f =
0.98$.  Although we do not have the needed sensitivity to determine
the core covering factor for this component in the VLBA data, warm gas
in our Galaxy is distributed more widely and more uniformly than the
cold gas (Dickey \& Lockman 1990), so it seems unlikely that the gas
would have a lower core covering factor than the cold gas.  There is
the possibility that the warm gas covers one or both of the weak
extended radio lobes as well as the core (ie. that $f > 0.98$), but
given that $30\arcsec \approx 45 h_{75}^{-1}$ kpc at $z_1=0.0912$, it
seems unlikely.  The absorption limits against the extended lobes (W.
Lane, in preparation) and the $EW_{21}$ of the warm absorption in the
Arecibo spectrum rule out the possibility that the absorption covers
one of the lobes but not the core as well.

When the warm and cold component column densities are added together,
the total column density in 21cm absorption on this sightline is
\NHI$_\mathrm{21cm} \leq 1.48 \pm 0.24 \times 10^{21}$ cm$^{-2}$.
This is in remarkable agreement with the column density from the fit
to the DLA line.  

\section{DISCUSSION}

The existence of a second gas phase has often been suggested to
explain the large harmonic mean spin temperature values, typically
\Ts\ $\approx 1000$ K (cf. Carilli et al. 1996) found in redshifted
DLA/21cm absorbers.  If the gas on the sightline has two (or more)
temperature phases, then the $T_{\mathrm s}$ calculated by comparing
the 21cm and DLA absorption profiles will not be equal to the kinetic
temperature in either phase, but rather to a column-density weighted
harmonic mean of the temperature of each phase.  Thus a sightline with
mostly warm phase gas will have a higher calculated \Ts\ than one with
mostly or only cold phase gas.  The values found for \Ts\ can then
best be interpreted as an upper limit to the temperature of the cold
phase gas.  When this quantity is derived for redshifted systems, it
is usually a value somewhere between the measured temperatures of
stable cold and stable warm neutral gas in our own Galaxy.  In most
redshifted 21 cm absorbers, the absorption lines are broadened by bulk
kinematic motions of the gas, and the limit on the kinetic temperature
is higher than that of the \Ts.

This is the first redshifted 21cm absorber measured for which the
calculated thermal $T_{\mathrm k}$ which constrains the cold gas
temperature more tightly than the derived \Ts, and shows directly that
not all of the gas column density seen in the DLA line appears in the
cold components of the 21cm line, as we expect from the value of \Ts.
It suggests that some two-thirds of the column density on this
sightline, if not more, is contained in warm phase gas.  If the broad
shallow component we have detected in our Arecibo data were to be
resolved by more sensitive observations into a collection of shallow
narrow absorptions lines, then the conclusion that warm phase gas is
necessary to explain all of the DLA fitted neutral column density in
this system would still remain.  The integrated \NHI\ in such an
ensemble of little narrow cold components would be very small, and we
would still need to find the rest of the gas sensed by the DLA
observation.  The logical place for it to be ``hiding'' from the 21cm
absorption measurement would still be in a warm neutral component,
appearing as an even broader and shallower absorption feature.

This system offers an explanation for why some DLA absorbers which
fall in front of radio-bright QSOs do not show 21cm absorption.  There
is \NHI\ $\approx 10^{21}$ cm$^{-2}$ column of gas which was
``unseen'' in the 21cm spectrum until extremely sensitive observations
were made.  This amount is well over the canonical lower limit for DLA
systems, and its existence here suggests that some fraction of the DLA
systems may arise in entirely warm phase gas.  This idea is
strengthened by the lower limits found on \Ts\ for 21cm
non-detections, which are typically several times 10$^3$ K, suggestive
of either unstable or warm phase gas (WNM) (Carilli et al.  1996).
Given that the WNM is more widely and uniformly distributed than the
CNM in our own Galaxy (cf. Dickey \& Lockman 1990) and in observations
of local Dwarfs (Young \& Lo 1996,1997), it seems likely that the gas
responsible for any detection of DLA absorption would include a large
fraction of warm phase neutral gas.

In conclusion, we are able to derive not only a harmonic mean spin
temperature for the B0738+313 system, but also a thermal kinetic
temperature for the gas in the narrow-line absorbing clouds.  For the
first time in a redshifted \HI\ 21cm absorber, the kinetic temperature
derived from the velocity width of the absorption is observed to be
smaller than the derived spin temperature.  The neutral gas must be
split into warm and cold phases in order to account for this
discrepancy.  We find that the 21cm absorption spectrum is best fit by
three components; two narrow deep lines, assumed to arise in cold gas
with $T_{\mathrm k} \leq 300$ \& $105$ K, and a broad shallow
absorption feature which we identify with the warm phase gas at
$T_{\mathrm k} \leq 5050$ K.  Within the errors, all of the neutral
hydrogen column density seen in the DLA measurement is recovered by
these three 21cm absorption components.  This is the first detection
of warm phase gas in absorption in an extragalactic system, and the
highest redshift detection of warm neutral phase gas known.  It is
also the first time limits on temperature have been made in a
redshifted DLA system which show a two temperature distribution of the
neutral gas, comparable to that found in our own Galaxy.

\begin{acknowledgements} 
  
  We wish to acknowledge Karen O'Neill for her help in obtaining the
  Arecibo data, and R. Rutten for taking the WHT spectrum.  Support
  for W. Lane is provided by an Ubbo Emmius fellowship for graduate
  study at the Rijksuniversiteit Groningen.

\end{acknowledgements}

\begin{table}[hp]
\caption[]{Three Component Fit to the 21cm Absorption}
\begin{flushleft}
\begin{tabular}[h]{lccc}                                                  
\tableline\tableline   
Component:        &  1  & 2  & 3                \\
\tableline                                                          
Optical depth:  & 0.2462$\pm$0.0010  & 0.02527$\pm$0.0010  & 0.0063$\pm$0.0008\\
$\Delta$ V$_\mathrm{offset}$(km~s$^{-1}$) &  ~~  & 7.69$\pm$0.04 &   -1.59$\pm$0.66\\
FWHM(km~s$^{-1}$) & 3.687$\pm$0.019 & 2.18$\pm$0.11 &   15.2$\pm$1.4\\
$T_{\mathrm k}$(K) &  297$\pm$3  & 103$\pm$10  &   5046$\pm$953\\
$N_\mathrm{HI}(10^{20}$ cm$^{-2}$)  & $5.27\pm0.02$ & $0.11\pm0.01$ &$9.4\pm2.3$\\ 
\tableline    
\end{tabular}
\label{tab:21cm}
\end{flushleft}
\end{table}

\begin{figure}[htbp]
\centering
\leavevmode
\epsfxsize=1.0\columnwidth
\epsfbox{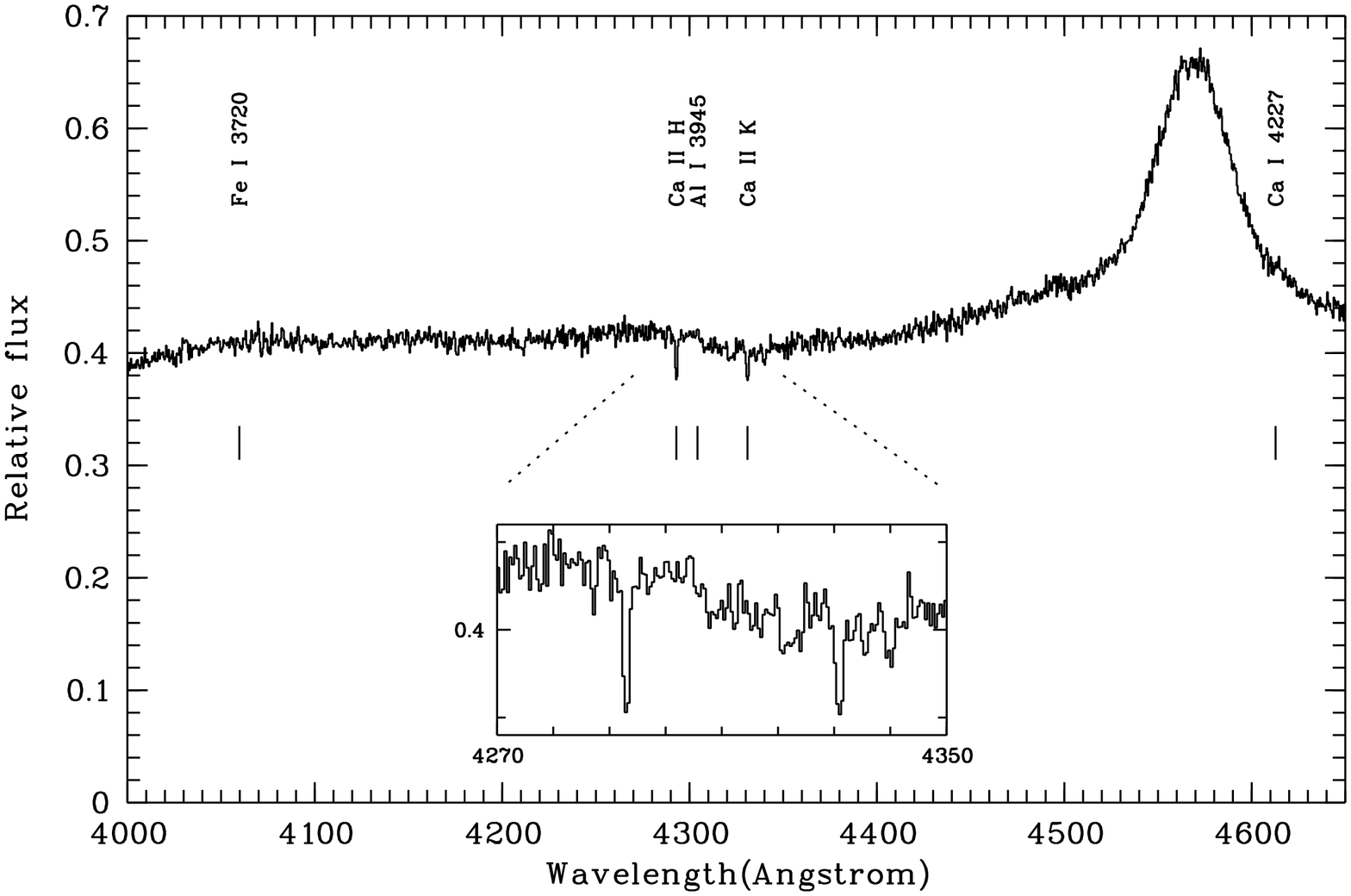}
\caption{WHT/ISIS spectrum of Q0738+313.  Ca {\sc ii} H and K
  absorption features at a redshift of $z=0.0910$ are marked.  No
  significant absorption is seen at the expected positions for Fe {\sc
    i} $\lambda 3270$, Al {\sc i} $\lambda3945$, or Ca {\sc i}
  $\lambda4227$.}
\end{figure}

\begin{figure}[ht]
\centering
\epsfxsize=1.0\columnwidth
\epsfbox{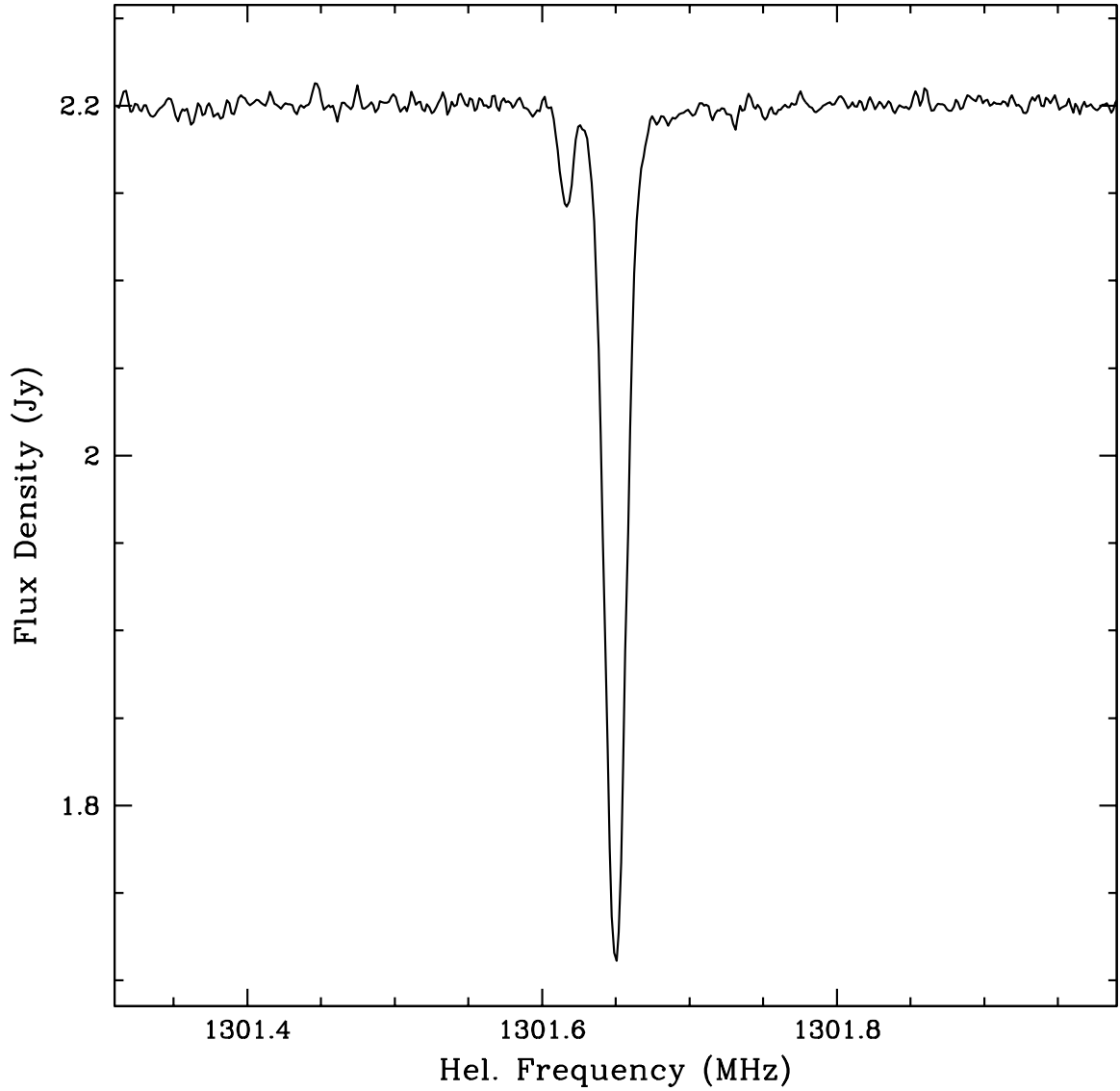}
\caption{Arecibo observations of the \HI\ 21cm line at $z=0.0912$
  towards the QSO B0738+313.  Channel spacing is 0.35 km s$^{-1}$,
  allowing this extremely narrow line to be resolved for the first
  time, and showing the second component clearly separated from the
  main component.}
\end{figure}

\begin{figure}[ht]
\centering
\leavevmode
\epsfxsize=1.0\columnwidth
\epsfbox{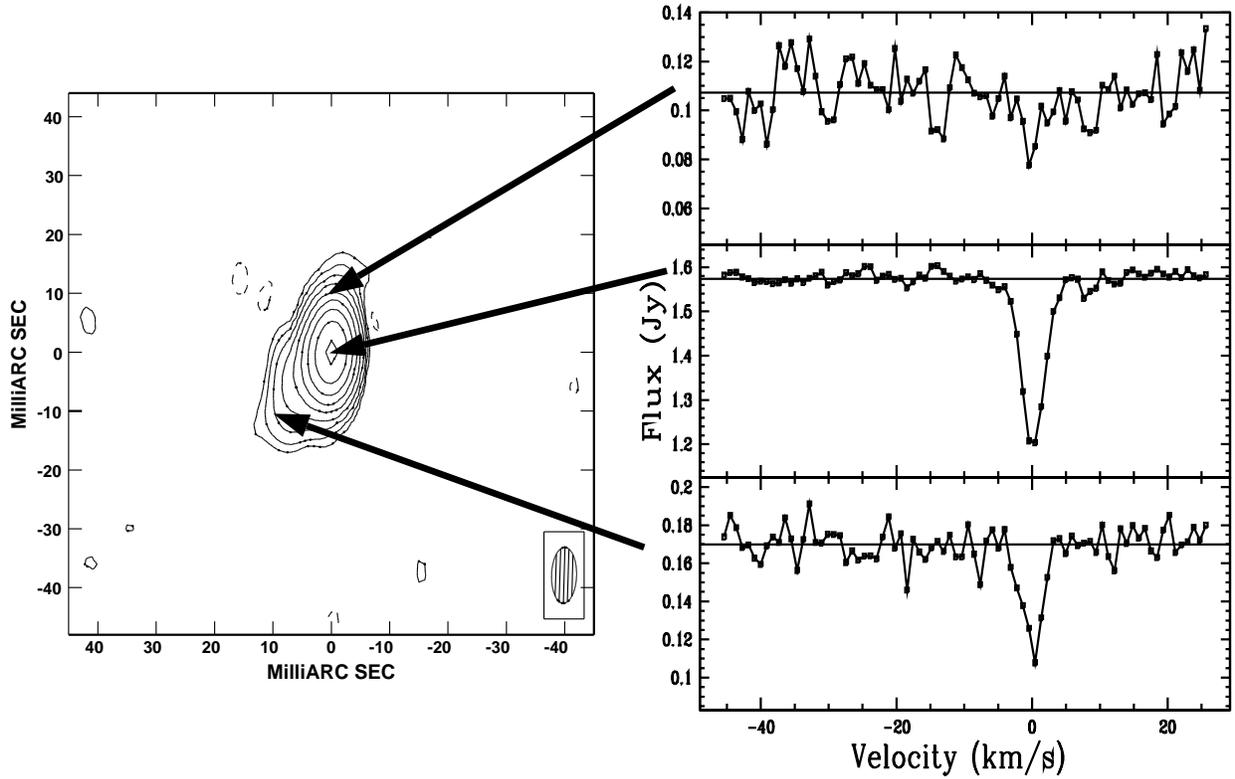}
\caption{Contour plot of the core of B0738+313 observed with the VLBA
  at 1302 MHz.  The quasar is lightly resolved showing a slight
  extension to the southeast.  Spectra have been extracted at the
  three positions indicated by the arrows.  The top spectrum is at
  $\delta x = 0, \delta y = 10$ mas, the center is at $\delta x = 0,
  \delta y = 0$, and the bottom at $\delta x = -10$ mas, $\delta y =
  -10$ mas.  At a redshift of $z=0.0912$, 10 mas corresponds to 15
  $h_{75}^{-1}$ pc.  The synthesized beam (FWHP) is roughly 5 mas by
  10 mas.  The velocity scale is centered with $v = 0$ at a
  heliocentric frequency of 1301.65 MHz.}
\end{figure}

\begin{figure}[ht]
\centering
\epsfxsize=1.0\columnwidth
\epsfbox{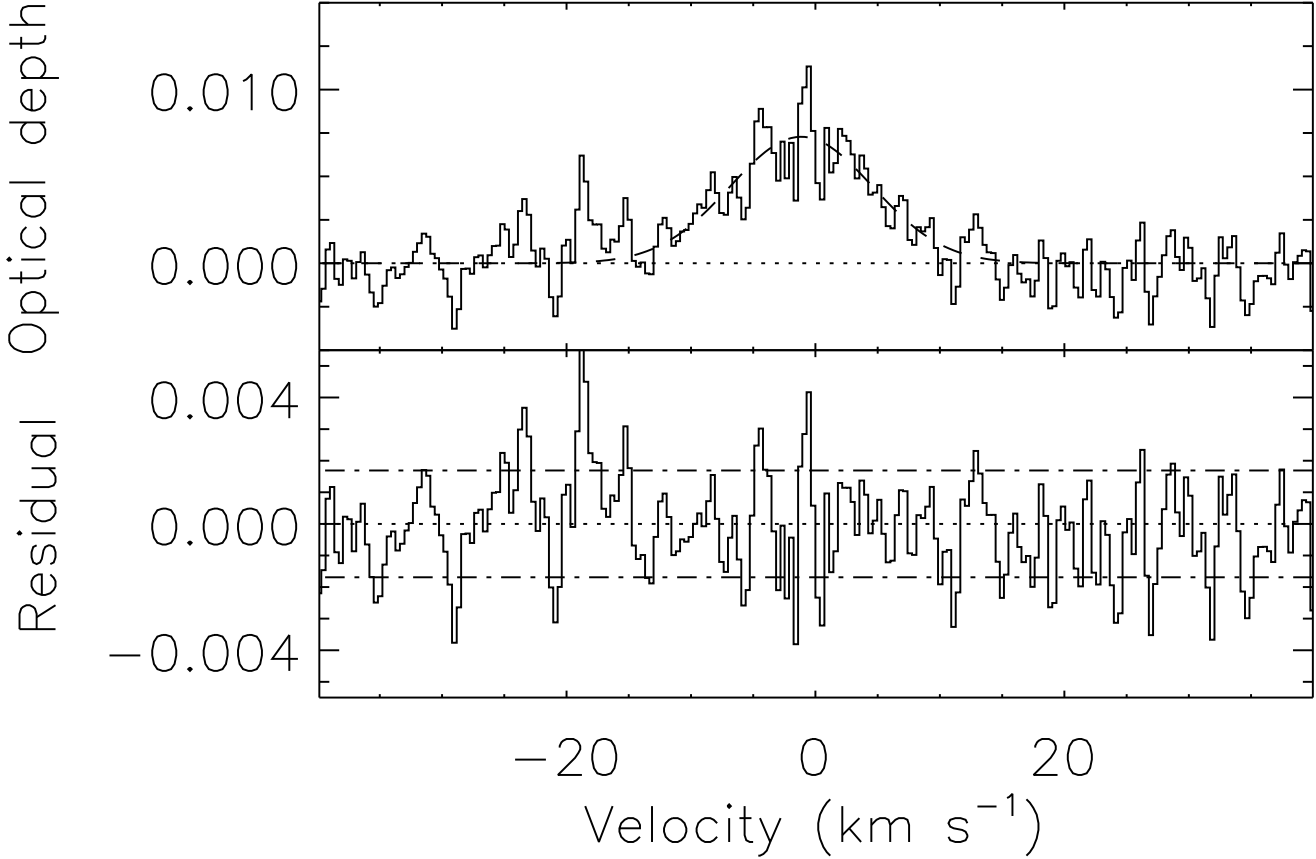}
\caption{Arecibo spectrum showing the warm component and its fit.  The
  velocity scale is set so that $v = 0$ corresponds to a redshift of
  $z_1 = 0.09123$.  The upper panel shows the spectrum after the
  Gaussian fits to the two cold component absorption lines have been
  subtracted.  The third warm component is clearly visible, and a
  dashed line shows the gaussian fit to it simultaneously with the two
  cold components.  The lower panel shows the residuals after
  subtracting all 3 components, with 1$\sigma$ error levels marked by
  the dash-dot lines.}
\end{figure}

\end{document}